\documentclass[runningheads,a4paper]{llncs}

\usepackage{amssymb}
\usepackage{graphicx}

\usepackage{url}
\newcommand{\keywords}[1]{\par\addvspace\baselineskip
\noindent\keywordname\enspace\ignorespaces#1}

\begin{document}

\mainmatter  

\title{Students' Comparison of Their Trigonometric Answers with the Answers of a Computer Algebra System\thanks{This research was funded by Estonian target funding project SF0180008s12. The final publication is available at http://link.springer.com.}}

\titlerunning{Students' Comparison of Their Answers with the Answers of a CAS}

%
%
\author{Eno Tonisson}%
\authorrunning{Students' Comparison of Their Answers with the Answers of a CAS}

\institute{Institute of Computer Science, University of Tartu, \\2 J. Liivi Tartu, 50409, Estonia \\
\email{eno.tonisson@ut.ee}}

\toctitle{}
\tocauthor{}
\maketitle

\begin{abstract}
Comparison of answers offered by a computer algebra system (CAS) with answers derived by a student without a CAS is relevant, for instance, in the context of computer-aided assessment (CAA). The issues of identity, equivalence and correctness emerge in different ways and are important for CAA. These issues are also interesting if a student is charged with the task of comparing the answers. What will happen when students themselves are encouraged to analyse differences, equivalence and correctness of their own answers and CAS answers? What differences do they notice foremost? Would they recognise equivalence/non-equivalence? How do they explain equivalence/non-equivalence?
The paper discusses these questions on the basis of lessons where the students solved trigonometric equations. Ten equations were chosen with the aim to ensure that the expected school answer and the CAS answer would differ in various ways. Three of them are discussed more thoroughly in this paper.

\end{abstract}

\section{Introduction}

The answers offered by a computer algebra system (CAS) are evaluated (in literature) mainly from a professional user's point of view. CAS users could compare their (or others') answers with the answers of a CAS for various reasons. For example, a mathematical researcher could use a CAS to confirm hand-derived solutions (see \cite{Bunt13}). Some mathematicians use a CAS in order to check the solutions of students (see \cite{Marshall12}).

There are also broader overviews, for example, in \cite{Wester99} where hundreds of answers are evaluated in case of several CAS.  A comparison of answers offered by a computer algebra system with answers derived by a student could be also used. One fruitful area is computer-aided assessment (CAA) where a student's answer is assessed automatically with the help of ("invisible") CAS. The capability of a CAA system depends on the capability of the CAS. It is necessary for comparison and assessment to explore the issues of identity, equivalence and correctness, and not only in the sense of classical mathematics (see \cite{Bradford09} and \cite{Bradford10}).

Further interesting issues arise if a student (not a computer program) is charged with the task of comparing the answers. This paper focuses on the following questions: What will happen when students themselves are encouraged to analyse differences, equivalence and correctness of their own answers and CAS answers? What differences do they notice foremost? Would they recognise equivalence/non-equivalence? How do they explain equivalence/non-equivalence? As different systems could present answers in different ways, a particular CAS was prescribed to initiate an "intrigue" and obtain information about the effect of different representations.

The paper is based on lessons where first-year students solved trigonometric equations. Solving trigonometric equations is an interesting topic in this context because of the variety of possible presentations of solutions, units of measurement, general and particular solutions, etc.  

The students worked in pairs and their discussions were audio-taped. The students had worksheets with equations and tasks (see Sect. 4). The order of solvable equations was prescribed and was different for different pairs. The students first solved an equation (correctly or not) without and then with a particular CAS. The systems used were Maxima \cite{Maxima}, Wiris \cite{Wiris}, and WolframAlpha \cite{WolframAlpha}. A specific CAS was prescribed for the equation to attain the expected difference between the students' answers and the CAS answer. 

Data of more than 100 instances of equation-solving (29 pairs of students) were collected. Three equations from ten are more interesting from the MKM point of view and they were chosen for deeper analyses in this paper (47 instances of equation-solving, 26 pairs of students).  

Before comparing their own answers with the answers of a CAS, students should read and understand CAS answers. Generally, the issue of readability of a CAS answer is multi-faceted. On the one hand, it is connected to sophisticated mathematical reasoning, for example, branch cut and cylindrical algebraic decomposition (CAD) (see \cite{Bradford02}, \cite{Phisanbut11}). On the other hand, sometimes a rather simple difference (for example, in notation) between a CAS answer and a school-like answer could be confusing for the student (see \cite{Drijvers02}). The issue of readability of CAS answers is important for this study. 
 
The broader perspectives of the topic could be described in teaching-learning context and in research context. The further purpose is to suggest a new method of using CAS for teaching and learning mathematics where students' discussion, critical thinking and deeper insight into important issues (such as equivalence) should be brought out. 
The readability of a CAS answer is as challenging as the learning of the CAS syntax. Moreover, the black box nature of a CAS reveals issues that can go as deep as university-grade mathematics (see \cite{Marshall12}). However, C. Buteau et al. note in \cite{Buteau10}: {\em Although practitioners have to deal with unusual or unexpected behaviour of CAS, this was occasionally shown to provide pedagogical opportunities.} 
A thorough study is needed for the method and this paper is a part of it. 
Besides direct teaching-learning context, perspectives in the research context are also important. Such an analysis of students' worksheets and discussions could provide new opportunities for studying their thinking and learning. A. Sfard (in \cite{Sfard08}) compares the possibility to analyse conversation even to a microscope that gives new power perspectives to 17th century scientists.

Accordingly, the study also includes a search for preliminary answers to questions about suitability of the method for teaching-learning and research.

Section 2 gives a brief overview of related works. The choice of equations is described in Sect. 3. An overview of the lessons is provided in Sect. 4. The main examples are discussed in Sect. 5, 6, 7. Section 8 concludes the paper.  

\section{Related Work}

This study is related to a number of different research areas. For example, comparisons of the different CAS, like \cite{Wester99}, \cite{Bernardin96} and \cite{Bernardin99} are notable. Such reviews do not focus on pedagogical aspects. However, M. Wester mentioned: {\em One could invoke mindset} ({\verb+Elementary_math_student+) {\em  to initially declare all variables to be real, make $\sqrt{-1}$ undefined, etc., for example} \cite{Wester99}. The adequacy of CAS answers is under consideration in \cite{Stoutemyer91} and \cite{Drijvers02}, for instance. The paper \cite{Drijvers02}, based on the experiments of P. Drijvers, is focused on parameters, but also defines a more universal list of obstacles. In addition, he suggests that an obstacle could be an opportunity. The CAS answers are observed from a school-oriented point of view in \cite{Tonisson08} and \cite{Tonisson11}.

The papers \cite{Bradford09} and \cite{Bradford10} were written from the background of CAA. CAA systems could be connected with a CAS. For example, the STACK system uses Maxima \cite{Sangwin07}. The issues of identity, equivalence and correctness are very important. They help to distinguish between mathematical, pedagogical and aesthetic correctness. Their study is focused on automated assessment. In our study the students were charged with the task of comparing the answers themselves. The issue of the "right answer" is also very important in this case. 
Our study is also related to the analysis of discourses, audio recordings, etc., but these topics are too far removed from the main focus of this paper. Furthermore, we do not deal here with the theoretical background of checking equivalence (for example \cite{Richardson68}) where trigonometry has a somewhat problematic status. The topic of trigonometric equations is considered in the next section.

\section{Choice of Equations and CAS}

Our study is focused on trigonometric equations because of the variety of their answers. It is quite usual for a trigonometric equation to have several reasonable representations of the correct answer. Different solution strategies may lead to different-looking but still equivalent answers. A classroom discourse in case of the equation 
\begin{equation}
2+ \cos ^2 2x = (2-\sin ^2x)^2
\end{equation}
is presented in  \cite{Abramovich05}. Four different answers were under consideration.

The variety of answers is actually even larger as the circumstances involved go beyond a pure solving strategy. For example, one could prefer radians or degrees. General solutions can be sought in some and particular solutions in other instances. Some basic formulae could be slightly different in different regions. For instance, the solution for  $\sin x = m$, could be expressed as 
\begin{equation}
\begin{array}{l} x = \arcsin m + 2 n \pi, \ n\in \mathbb{Z} \\ x = \pi - \arcsin m + 2 n \pi,\ n\in \mathbb{Z}\end{array}
\end{equation}	
or (as in Estonian textbooks, for example)
\begin{equation} \label{eq:mn}
x=(-1)^n \arcsin m + n \pi,\ n\in \mathbb{Z} \enspace .
\end{equation}

If we use a CAS, the variety is likely to increase because of the peculiarities of the CAS. For example, some notation issues could arise. Different treatments of the (default) number domain can also have an impact. Nevertheless, the issues of general and particular solutions or "regional" differences could be relevant.  
In this study, 10 equations were chosen for the class. Some of them were from regular school textbooks, others from books where trigonometry is handled at a somewhat advanced level. We analyse three of them in more depth in the paper. These equations seemed to be more suitable for this research track, as the focus is primarily on different representations of the answers and not so much on extraneous roots, complex domain, etc., (like in case of some other equations).
The equations were chosen to attain a specific type of difference between the expected answer of the students and the answer of the particular CAS. (Actually, as students solved the equations themselves, they also made mistakes and the comparison was made between their actual answers and the CAS answers. It is more thoroughly explained in the next section.)
	
The first example is the equation $\sin(4x+2)=\frac{\sqrt{3}}{2}$ , where the students use Formula \ref{eq:mn} (as taught in Estonian schools) but WolframAlpha expresses series separately (see Sect. 5). (The issue of branches in CAS is also discussed in \cite{Tonisson07}). The second example is $\tan ^3 x = \tan x$, where students give general solutions but Wiris gives particular solutions (see Sect. 6). The third example is $\cos\left( x-\frac{\pi}{6} \right) =0.5$, where Maxima uses its own notation with \verb+union+ and \verb+%z+ 
(see Sect. 7).
The other equations with more specific nuances (extraneous roots, issues of domain, indeterminacy, etc.) are not discussed in this paper but are listed for the sake of completeness: $2 \sin 2x \cos 2z + \cos 2x =0$, $\displaystyle \frac{\tan ^2 x}{\tan x}=0$, $\tan(x+\displaystyle \frac{\pi}{4})=2\cot x -1 $, $2\cos ^2 x + 4 \cos x=3\sin^2 x $, $\sin x - \sin ^2 x = 1+\cos^2 x$, $\displaystyle \frac{1-\cos x}{\sin x}=0$, $1-\cos x = \sqrt{3} \sin x$. 

\section{In Class}

This section gives a brief overview of the lessons in the course "Elementary mathematics", which is a somewhat repetitious course of school mathematics for the first-year university students. The students had quite diverging skill levels in solving trigonometric equations. As the advanced students were dismissed from the course (on the basis of a preliminary test), the proportion of wrong answers probably increased. 
The students had very few experiences with CAS. CAS were not used in other lessons of the course. 

The lesson in question was taught by the author (who was not a regular teacher of the course). The lesson lasted for 90 minutes and consisted of an introduction, a period of equation-solving (ca 70 minutes), and closing (saving and copying data). The introduction gave an overview of the lesson, the aims of the study, etc. The computer algebra systems were not specially introduced but the students were warned that the answers of a CAS could differ from human answers and could also be incorrect. The types of possible differences were not explained.
The order of solvable equations was prescribed and was different for different pairs of students in order to collect data about different equations. The students solved the equations in pairs and the discussions were audio-taped in order to obtain a deeper overview beyond the notes on paper. The students first had to solve the trigonometric equations by themselves and then with a particular CAS. They were encouraged to analyse differences, equivalence and correctness of their own answers and CAS answers. The worksheet included the following tasks (in the case of the first example):

\begin{itemize}
\itemsep8pt\relax

\item Solve an equation $\sin(4x+2)=\frac{\sqrt{3}}{2}$ (without the computer at first).
\item How confident are you in the correctness of your answer?
\item Solve the equation with the CAS WolframAlfa using the command solve.
\begin{figure}
\centering
\includegraphics[width=10cm]{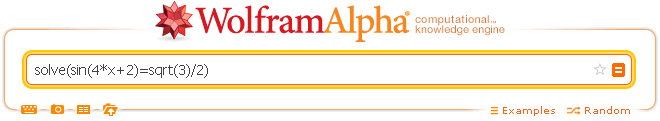}
\caption{WolframAlpha input}
\label{fig:wain}
\end{figure} 
\item How unexpected is the CAS answer at first view? 
\item Analyse the accordance of your answer with the CAS answer!
If you want to complement/correct your solution, please use the green pen.
\item What are the differences between your answer and the CAS answer?
\item How are your answer and the CAS answer related (analyse equivalence/non-equivalence, particular solutions/general solutions)?
\item Rate the correctness of your (possibly corrected) answer.
\item Rate the correctness of the CAS answer.
\end{itemize}

\noindent  
Some of the issues are discussed thoroughly in this paper, others are mentioned only in the conclusion where further work is described.

The student papers and audio-tapes were analysed and the results in case of three examples are presented in the next sections. Each presentation begins with a brief introduction of the example, including reasons for selecting the example, a possible school answer, and a snapshot of the CAS answer. Next, the equivalence/non-equivalence of the students' answers with the CAS answers is discussed. It is based on mathematical reasoning by the author ({\em Math.} in tables). The second dimension is the students' opinion about the equivalence/non-equivalence that is based on an analysis of paper and audio data ({\em Stud.} in tables). The tables are also presented. The discussion concludes with some pedagogical comments.

\section{Different Forms of General Solution}

\begin{figure}
\centering
\includegraphics[width=12cm]{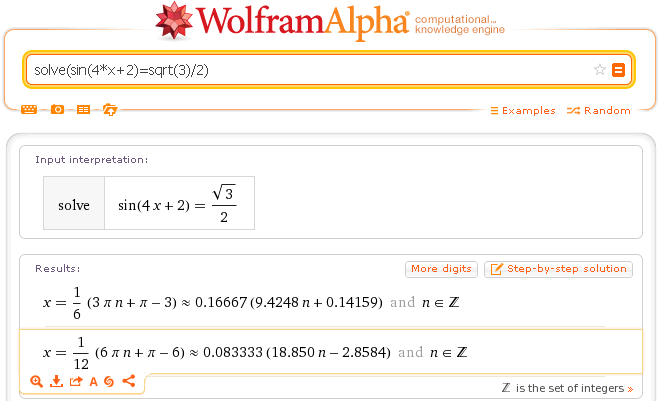}
\caption{$\sin(4x+2)=\frac{\sqrt{3}}{2}$. WolframAlpha}
\label{fig:waout2}
\end{figure}

The first example is the equation where the CAS answer is particularly unexpected for those who use the $(-1)^n$ formula (Formula \ref{eq:mn}) in case of $\sin x = m$ (as is common for Estonian students). The possible school answer for the equation 
\begin{equation}
 \sin(4x+2)=\frac{\sqrt{3}}{2}
\end{equation}
is
\begin{equation}
x=-\frac{1}{2}+(-1)^n\frac{\pi}{12}+\frac{n\pi}{4},\ n\in \mathbb{Z} \enspace .
\end{equation}

WolframAlpha gives two series of solutions (see Fig.~\ref{fig:waout2}). The answers are actually equivalent. The students did not receive any specific information about the CAS answer.

As our textbooks and teachers use mainly the $(-1)^n$ form, the students' answers and the CAS answer seemed quite different at least for this reason. (Twelve pairs (of 17) used $(-1)^n$ form, 4 gave particular solution. One pair gave initially particular solution and after correction $(-1)^n$ form.) 

As several pairs made mistakes, one could count 11 cases of equivalence with the CAS answer and 6 cases non-equivalence.    
Four pairs (of equivalent cases) used both degrees and radians in the same answer, for example:
\begin{equation}
x=15^\circ \cdot (-1)^n+45^\circ \cdot n -\frac{1}{2},\ n\in \mathbb{Z} \enspace .
\end{equation}

Our main focus in the paper is to observe how students compare their own and CAS answers. In many cases, their opinion about the equivalence is ascertainable, sometimes not. The results are presented in Table 1.

\begin{table} \label{tab:ex1}
\caption{$\sin(4x+2)=\frac{\sqrt{3}}{2}$. Equivalence/non-equivalence}
\begin{center}
\begin{tabular}{|l|c|c|c|}
\hline

         ~              & Stud. Equivalent & Stud. Non-equivalent & Abstruse \\ \hline
         Math. Equivalent     & 4          & 5              & 2        \\ \hline
         Math. Non-equivalent & 3          & 3              & ~        \\ \hline

\end{tabular}
\end{center}
\end{table}
The depth of discussions about the comparison varied between the student pairs. For example, 3 pairs identified actual equivalence through reasonable discussion, while one pair simply presumed it. There were also 3 pairs whose answer was not equivalent with the CAS answer, but they counted them as equivalent without any real discussion. Seven pairs did not recognize that the answers were equivalent. Mainly, they did not grasp that $n$ in their answer (like in Formula~\ref{eq:mn}) and $n$ in the CAS answer (see Fig.~\ref{fig:waout2}) was not the same. This points to an automated (and correct) habit of solving the algorithm of trigonometric equation without exhaustive understanding of the solution. Three pairs identified the non-equivalence of their answer and the CAS answer. Their answers were remarkably different from the CAS answer. 

It seems that the different representations of the same answer, like in this example, could initiate instructive discussion. It could also point to a possible superficial treatment of the fairly important issue of the meaning of $n$. A simpler equation, like $\sin4x=\frac{\sqrt{3}}{2}$ , could probably be a more straightforward means for clarifying the phenomenon. The example is suitable if the students use the $(-1)^n$ formula. 
This is also an issue of different traditions. For example, it is usual to find solutions such as 
\begin{equation}
x=(-1)^n\frac{\pi}{6}+n\pi,\ n\in \mathbb{Z}
\end{equation}
(being the solution of $\sin x = \frac{1}{2}$) in the textbooks of some countries, such as Estonia, whereas we are aware that many others do not.

\section{If the CAS Gives Only Particular Solutions}

As the second example, we have chosen a situation where a CAS gives only particular solutions, but the students were asked to present general solutions. Wiris has its own rules for the presentation of solutions to trigonometric equations. In case of $\sin x = a$, for example, $\arcsin a$ and $\pi-\arcsin a$ are presented. 

The students should frame the CAS solutions up to their own general solutions. In case of the equation
\begin{equation}
\tan ^3 x = \tan x
\end{equation}
the human answer could be 
\begin{equation}
\begin{array}{l} x = n\pi, n\in \mathbb{Z} \\ x = \pm \displaystyle \frac{\pi}{4} + n\pi, n\in \mathbb{Z} \end{array}
\end{equation}
or
\begin{equation}
\begin{array}{l} x = n\pi, n\in \mathbb{Z} \\ x = \displaystyle \frac{\pi}{4} + n\pi, n\in \mathbb{Z} \\ x = -\displaystyle \frac{\pi}{4} + n\pi, n\in \mathbb{Z} \enspace .\end{array}
\end{equation}

Wiris gives the particular solutions (Fig.~\ref{fig:wiout3}).
\begin{figure}
\centering
\includegraphics[width=10cm]{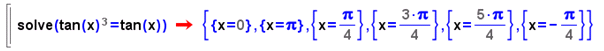}
\caption{$\tan ^3 x = \tan x$. Wiris}
\label{fig:wiout3}
\end{figure}

We count these answers as equivalent in the sense that all series are presented by 2 instances. Certainly, $n\pi$ and $\{0; \pi\}$ are not equivalent in the usual mathematical sense. The order of solutions is quite confusing as instances of the series of solutions are not always side by side (for example, $\frac{\pi}{4}$ and $\frac{3\pi}{4}$ are not from same "club"). The students did not receive any specific information about the CAS answer.
Many of the student pairs (9 of 14) gave the right answer and they also figured out (after smaller or larger effort and discussion) the relationship between their and CAS answer (see Table 2). One pair could not frame $\pi$, $\frac{3\pi}{4}$ and $\frac{5\pi}{4}$ up to their right answer. Again, the meaning of $n$ in the formula seemed to be incoherent for them. The cases where students omitted some solutions were very interesting. One such pair corrected their mistake and finally found the right answer. They added to 
\begin{equation}
\begin{array}{l} \pi + \pi n \\ \displaystyle \frac{\pi}{4} + \pi n \end{array}
\end{equation}
missing
\begin{equation}
 -\frac{\pi}{4} + \pi n \enspace .
\end{equation}

We do not focus on emotions in this paper but their joy after the correction was remarkable. The other pair (initially only $n\pi$ solution) had a member who already diagnosed their mistake. 
The third pair did not analyse the CAS solutions thoroughly enough and did not notice that their answer was incomplete. It is impossible to give a thorough overview of the discussion of the pair that got an incomplete answer and also considered it as non-equivalent with the CAS answer, as their discussion was very laconic.  
\begin{table} \label{tab:ex2}
\caption{$\tan ^3 x = \tan x$. Equivalence/non-equivalence}
\begin{center}
\begin{tabular}{|l|c|c|}
\hline

         ~              & Stud. Equivalent & Stud. Non-equivalent \\ \hline
         Math. Equivalent     & 9          & 1              \\ \hline
         Math. Non-equivalent & 2          & 1 \\ \hline
         Non-equivalent $\rightarrow$  Equivalent  &1& ~        \\ \hline

\end{tabular}
\end{center}
\end{table}
It seems that the representation of the answer is generally accomplishable in this case. The possible corrective virtue is also notable. The standard of representation of answers to trigonometric equations could provide more instructive examples, as the choice of a particular solutions is not always as transparent.

\section{Unusual Form of Arbitrary Integer}

The third example is related to CAS notation. The CAS answer is actually very similar to a normal human answer but with some CAS-specific peculiarity.  The human answer to the equation
\begin{equation}
\cos\left( x-\frac{\pi}{6} \right) =0.5
\end{equation}
could be
\begin{equation}
\begin{array}{l} x = - \displaystyle \frac{\pi}{6} + 2n\pi, n\in \mathbb{Z} \\ x = \displaystyle \frac{\pi}{2} + 2n\pi, n\in \mathbb{Z} \enspace .\end{array}
\end{equation}
Maxima gives the same answer in a somewhat distinctive way (Fig.~\ref{fig:max}). The package} \verb+to_poly_solve+ is used for solving trigonometric equations. We cite the Maxima manual for clarity: {\em Especially for trigonometric equations, the solver sometimes needs to introduce an arbitrary integer. These arbitrary integers have the form \verb+%zXXX+, where \verb+XXX+ is an integer} \cite{maximaman}.
 
\begin{figure}
\centering
\includegraphics[width=6.5cm]{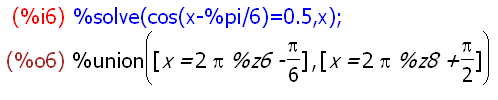}
\caption{$\cos\left( x-\frac{\pi}{6} \right) =0.5$. Maxima}
\label{fig:max}
\end{figure}
The meaning of \verb+%z+ was also an important issue for solving the equation with Maxima. The students did not receive any specific information about the CAS answer, but they had additional brief paper manuals (3 pages) on using different CAS where \verb+%z+ was explained. Only two pairs found the info about \verb+%z+ from this manual. Almost all pairs mentioned the \verb+%z+ as a remarkable difference from their own answer. An explanation was given if the students asked about it. Nevertheless, two pairs remained confused and could not understand the CAS answer. The meaning of such a notation could be more clearly indicated in the CAS user-interface. For example, tooltips could be used.
Eight pairs (of 16) got the right answer (see Table 3). Five of these pairs quite easily found the CAS answer to be equivalent. Three pairs had an answer equivalent with the CAS answer but their opinion about equivalence was abstruse. One of these pairs could not understand the CAS answer because of \verb+%z+. The second pair did not observe the CAS answer sufficiently and did not notice the relation between the CAS answer and their own (not fully simplified) answer. The third pair's discussion was too laconic.     
One pair corrected their mistake and finally found the right answer, from 
\begin{equation}
\begin{array}{c} \dots \\
x-30^\circ = \displaystyle \arccos\frac{1}{2} + 2\pi n \\
\dots 
\end{array}
\end{equation}
to
\begin{equation}
\begin{array}{c}
\dots \\
x-30^\circ = \pm \displaystyle \arccos\frac{1}{2} + 2\pi n \\
\dots \enspace .
\end{array}
\end{equation}

Three pairs saw equivalence that really did not exist. There were also four pairs who considered their wrong answers as non-equivalent with the CAS answer. One of these pairs could not understand the meaning of \verb+%z+ correctly. Two pairs tried to find their mistakes, one pair had evidently a different answer.  
\begin{table}
\caption{$\cos\left( x-\frac{\pi}{6} \right) =0.5$. Equivalence/non-equivalence} \label{tab:ex3}
\begin{center}
\begin{tabular}{|l|c|c|c|}
\hline

         ~              & Stud. Equivalent & Stud. Non-equivalent & Abstruse \\ \hline
         Math. Equivalent     & 5          &               & 3        \\ \hline
         Math. Non-equivalent & 3          & 4              &         \\ \hline
		 Non-equivalent $\rightarrow$ Equivalent  & 1  &      &      \\ \hline
		
\end{tabular}
\end{center}
\end{table}

It seems that the different notation can cause major trouble for some people, while it can be easily acceptable for others. It should be mentioned that the students used Maxima for the first time and many issues would probably be resolved in the course of further use.

\section{Conclusion}

The study focused on a lesson where students solved trigonometric equations at first without a CAS and then with a CAS. The main task was to compare the answers. Recognition of equivalence can be a difficult task for students. Even those students who solve trigonometric equations quickly and correctly can find it hard to correctly compare their answer with the CAS answer. The three examples presented in the paper helped to highlight different aspects of this situation.

If we look at the findings in Sections 5, 6 and 7, it is possible to single out the cases where students identified the equivalence/non-equivalence of their answer and the CAS answer adequately. The proportions of these cases are presented in Table 4. The cases where the non-equivalent answer was changed to equivalent in the light of the CAS answer are also included.

\begin{table}
\caption{Adequate identification of equivalence/non-equivalence} \label{tab:con}
\begin{center}
{
\renewcommand{\arraystretch}{1.5}
\begin{tabular}{|l|c|c|c|}
\hline

         Section              & Equation & \multicolumn{2}{c|}{Adequate identification} \\ \hline
         Different Forms of General Solution     & $\sin(4x+2)=\frac{\sqrt{3}}{2}$ &       7 (of 17)        & 41\%        \\ \hline
         If CAS gives only particular solutions  & $\tan ^3 x = \tan x$   &     11 (of 14)          & 79\%        \\ \hline
		 Unusual Form of Arbitrary Integer  & $\cos\left( x-\frac{\pi}{6} \right) =0.5$  &   10 (of 16)   &   62\%   \\ \hline
		
\end{tabular}
}
\end{center}
\end{table}

There seem to be different "hindrances" to identification of equivalence/non-equivalence in case of different equations. Probably, the proportion of adequate identifications of equivalence/non-equivalence could be increased by drawing special attention to the problematic issues before solving or in the worksheets. It is important to decide what issues are adequate and useful for the students. For example, the meaning of $n$ in the answers of trigonometric equations is relevant and useful for mathematical insight. (The $\%z6$ topic is also connected to this issue.) It is probably possible to increase the proportion of students' answers that are equivalent to CAS answers. For example, it is possible to use simpler equations or give more hints about the solution. Principally, it is possible to give whole solutions but then the students would have a weaker connection with the exercise.

These questions could be studied in further experiments. Actually, there are various ideas for further work. Data from the lessons (see worksheet in the Sect. 4) include information about students' pen and paper solutions, confidence, rating of correctness of their answers and CAS answers, etc. The data could be analysed with the topic discussed in this paper. Of course, the seven equations not discussed in this paper should be included in the study. It is notable that some of these equations have CAS answers that are non-equivalent to school answers. 

As a teacher, the author could argue (so far without scientific proof) that the lessons were successful. (It was also confirmed by the actual teachers of these groups.) It seems that the task of comparing their own answers and CAS answers was interesting to the students. Generally, they became accustomed to the style of the lesson and actively discussed the topic of trigonometry. The method seems to be fruitful in research context as well. The paper data and audio-tapes complement each other and give a good overview about the students' activities during the solving process. 

Coming back to the issue of readability of the answer, it should be mentioned that $\%z6$ form could be confusing for some students, but it seems to be easily explainable. However, the change of $\%z6$ form could be considered as a possible suggestion to CAS developers. In addition, it is possible to improve the order of particular solutions in Wiris. 

It would be quite useful if a CAS would have the possibility to choose a mode according to a particular style of presenting the answers. For example, one could choose whether a general solution would be in $(-1)^n$ form or in the form of two series. On the one hand, it is good if the CAS answers are very school-like. On the other hand, the moderate difference between the students' and CAS answer could also be challenging and useful. Specification of such moderation is one of the most challenging issues of further work. It opens more questions. For example, how would such a specification look like? Would it be possible to work out indicators that qualify the type of answers?   

One could even say that having different answers compared to school solutions is a part of the charm of CAS. It is possible to propose various lesson scenarios other than those used in the lessons considered here. A discussion where all students would participate could be very useful. The discussion could take place during the same lesson after solving and comparing, but it is also possible to arrange the concluding discussion during the next lesson. In any case, the concluding part is desirable, as students need feedback. 

It is also possible to direct students to use CAS tools in the comparison of answers. For example, they could try to substitute a solution into the equation, simplify the difference of answers with the help of the CAS. Of course, it is possible that students compare their own answers with CAS answers as they did in these lessons. Another possible task for students could be a comparison of the answers of different computer algebra systems. In addition, one and the same CAS could offer different answers with different commands or assumptions and these answers could be also compared.

We can conclude that the method of asking students to compare their own answers with CAS answers seems to have potential in the context of learning as well as research, but further work is certainly needed. This style of comparison could contribute of the usage of computer-based tools for doing mathematics in different ways. On the one hand, the students see that calculations can be performed faster and easier. On the other hand, one should understand that evaluation of a CAS answer may not be so fast and easy. The abilities of critical thinking (particularly, with respect to computer algebra systems) are likely to be developed by the exercises.
\\


\end{document}